%% file: ms.tex
\documentclass[5p,longtitle,preprint]{elsarticle}

\usepackage{hyperref}
\usepackage{graphicx}  
\usepackage{dcolumn}   
\usepackage{bm}        
\usepackage{amsmath}   
\usepackage{amssymb}   
\usepackage{verbatim}
\usepackage{color}
\usepackage{textpos}

\journal{Physics Letters B}

\begin{document}
\sloppypar
\begin{frontmatter}
\title{Measurement of the single-spin asymmetry $A_y^0$ in quasi-elastic $^3$He$^\uparrow$($e,e'n$) scattering at $0.4 < Q^2 < 1.0 \mathrm{~GeV}/c^2$}

\input author_list_plb.tex

\newpageafter{author}

\begin{abstract}
Due to the lack of free neutron targets, studies of the structure of the neutron are typically made by scattering electrons from either $^2$H or $^3$He targets. 
In order to extract useful neutron information from a $^3$He target, one must understand how the neutron in a $^3$He system differs from a free neutron by taking into account nuclear effects such as final state interactions and meson exchange currents. The target single spin asymmetry $A_y^0$ is an ideal probe of such effects, as any deviation from zero indicates effects beyond plane wave impulse approximation. 
New measurements of the target single spin asymmetry $A_y^0$ at $Q^2$ of 0.46 and 0.96~(GeV/$c)^2$ were made at Jefferson Lab using the quasi-elastic $^3\mathrm{He}^{\uparrow}(e,e'n)$ reaction.  Our measured asymmetry decreases rapidly, from $>20\%$ at $Q^2=0.46$~(GeV/$c)^2$ to nearly zero at $Q^2=0.96$~(GeV$/c)^2$, demonstrating the fall-off of the reaction mechanism effects as $Q^2$ increases. We also observed a small $\epsilon$-dependent increase in $A_y^0$ compared to previous measurements, particularly at moderate $Q^2$. This indicates that upcoming high $Q^2$ measurements from the Jefferson Lab 12~GeV program can cleanly probe neutron structure from polarized $^3$He using plane wave impulse approximation. 
\end{abstract}


\end{frontmatter}

One of the fundamental goals of nuclear physics is to understand the structure and behavior of strongly interacting matter in terms of its basic quark and gluon constituents. Understanding the internal structure of nucleons is an important step towards this goal. Scattering electrons from light nuclei has been a proven method to probe these interactions \cite{Perdrisat:2006hj}.
While the structure of the proton is readily accessed by direct scattering of electrons on hydrogen targets, this technique cannot be used for neutrons since free neutron targets do not exist. 
Instead, scattering on particular nuclei is exploited, e.g. on $^2$H by virtue of its weak proton-neutron binding or $^3$He due to its spin properties being largely governed by the neutron \cite{Schulze:1992mb}. 
In order to extract the properties of the neutron from such studies, 
nuclear effects must be accurately taken into account. This drives the need to measure observables sensitive to such effects.

Assumptions made in the nuclear models can have a large effect on the extraction of the neutron form factors. 
In the late 1990s, there was a discrepancy between extractions of the electric form factor of the neutron, $G_E^n$, using the plane wave impulse approximation (PWIA) applied to data by electron scattering from deuterium \cite{Herberg:1999ud, Ostrick:1999xa} and $^3$He \cite{Meyerhoff:1994ev, Becker:1999tw, Rohe:1999sh}. 
This discrepancy was largely removed when full Faddeev calculations were used to extract the form factor instead of PWIA \cite{Golak:2000nt}. 
These calculations accounted for nuclear effects such as final state interactions (FSI) and meson exchange currents (MEC), which are ignored in PWIA.

The target single-spin asymmetry obtained by scattering electrons from a target polarized in two opposite directions transverse to the incoming electrons, $A_y^0$, is sensitive to these higher-order effects. This asymmetry is defined as
\begin{equation}
	A_y^0 = \frac{1}{P_t}\frac{N_{\uparrow}-N_{\downarrow}}{N_{\uparrow}+N_{\downarrow}},
\end{equation}
where $P_t$ is the polarization of the target and $N_{\uparrow}~(N_{\downarrow})$ is the number of normalized $^3$He$^{\uparrow}$($e,e'n$) events when the target is polarized parallel (anti-parallel) to the normal of the incoming electron beam.
In PWIA, this asymmetry is exactly zero \cite{Laget:1991pb}. 
Early predictions expected contributions from FSI and MEC to be large at low negative four-momentum transfer squared ($Q^2$) until dropping off at $Q^2$ of about 0.2~(GeV/$c$)$^2$ \cite{Laget:1991pb}. 
The first experimental test of $A_y^0$ done at NIKHEF showed this asymmetry to be 5.9$\sigma$ larger than expected \cite{Poolman:1999uf}. 
Another measurement was later performed at MAMI, which extended the measured $Q^2$ range up to 0.67~(GeV/$c)^2$ \cite{Bermuth:2003qh} with the same conclusion. 
Using full Faddeev calculations that correctly incorporated the significant effects of FSI, the predictions of Golak $et~al.$ agreed with the observed asymmetries \cite{Golak:2001ge}.
 This measurement of  $A_y^0$ provides unprecedented precision and extends up to $Q^2$ of 0.96~(GeV/$c)^2$. It provides new constraints on models used to extract neutron physics from electron scattering from $^3$He nuclei, and shows clear evidence of the dominance of nuclear effects across $Q^2$. 

\begin{figure}
	\centering
	\includegraphics{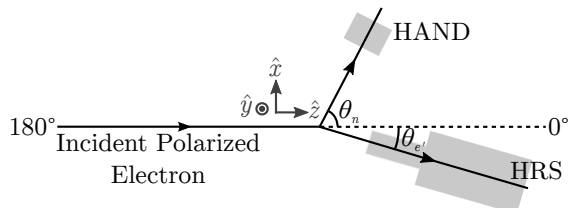}
	\caption [Hall A set-up] {Hall~A experimental set-up, where $\hat{y}$ is pointed along the vertical direction and $\hat{z}$ along the beam.
}
	\label{hall-angles}
\end{figure}

We report measurements on $A_y^0$ up to $Q^2$ of 0.96~(GeV/$c)^2$, performed at the Thomas Jefferson National Accelerator Facility (JLab) in Experimental Hall~A from April-May 2009. 
In the experiment, E08-005, a longitudinally-polarized electron beam with a current of 10 $\mu$A was incident on a polarized $^3$He gas cell. 
The beam helicity was flipped in a pseudorandom quad pattern every 33.3 ms  \cite{Chao:2011zz}. 
The target single-spin asymmetry measurement effectively assumed an unpolarized beam as events were summed over both helicity states. 
The small time frame of 33 ms between psuedorandom flips ensured than changes in luminosity between the two electron helicity states was negligible.
The beam, at energies of  2.4~GeV and 3.6~GeV, was incident on a 40-cm-long $^3$He cell that was polarized in the vertical $\hat{y}$ direction, as shown in Fig.~\ref{hall-angles}. Scattered electrons were detected in the high-resolution spectrometer (HRS) and knocked-out neutrons were detected using the Hall~A Neutron Detector (HAND)~\cite{Subedi2007,Long2012}. This experiment ran concurrently with multiple experiments that measured quasi-elastic structure on polarized $^3$He~\cite{Zhang:2015kna, Mihovilovic:2014gdi, Mihovilovic:2018fux, Sulkosky:2017prr}.

The $^3$He target was polarized through spin-exchange optical pumping (SEOP) 
\cite{Walker:1997zzc, PhysRevA.58.1412, PhysRevLett.91.123003,Liang01theepr, Incerti98thenmr}. 
An average target polarization of 
$51.4\pm0.8\pm4.6\%$ 
normal to the electron scattering plane was achieved. In order to reduce systematic uncertainties, the direction of the target spin was flipped by 180$^{\circ}$ every 20 minutes throughout the experiment, providing `up' and `down' target spin states. 

Electrons quasi-elastically scattered from the $^3$He nuclei were detected in a high-resolution spectrometer that consisted of three quadrupole magnets and one dipole magnet in a QQDQ optical chain, a pair of trigger scintillators, vertical wire drift chambers (VDC), a gas Cherenkov detector, and lead-glass calorimeters \cite{Alcorn:2004sb} for particle identification. The HRS calibration was identical to that in \cite{Sulkowski:2017}. 

Neutrons knocked out from $^3$He were detected by the Hall~A neutron detector, a non-standard piece of equipment used previously in a short-range correlation experiment~\cite{Subedi:2008zz}. It consisted of an array of plastic scintillators, each connected to photo-multiplier tubes on both ends. Electrons detected in the HRS acted as a trigger for HAND, which opened a timing window to detect correlated neutrons.

Since neutrons do not carry charge, they are not directly measured by the scintillator; however, they will produce a hadronic shower in the plastic scintillating detectors, which is detected. 
Since protons and neutrons are similar in mass, protons scattered from $^3$He will arrive at the detector at approximately the same time as neutrons. 
A proton will always deposit a signal in the 2-cm-thick veto bars whereas a neutron will most likely pass through the thin veto counter without interacting. However, for this experiment the thin veto layer was often flooded with accidentals that diluted the hadron time-of-flight spectrum and making proton rejection using the veto layer inefficient, particularly at low $Q^2$. In order to accurately identify neutrons, each layer of the HAND was used as a veto layer for the bars behind it, effectively creating a cascading veto layer. In addition at the highest $Q^2$ setting, a 9.08~cm thick wall made of 4~cm of iron casing surrounding 5.08~cm of lead was placed in front of the HAND that greatly reduced the number of background $\gamma$ events.

\begin{table}
\caption{\label{tab:qekinematics}Displayed are the kinematics settings for the $A_y^0$ measurements. Listed are the central four-momentum transfer, $\left<Q^2\right>$; beam energy, E$_0$; the HRS central angle, $\theta_{e'}$; the HRS central momentum, p$_0$; and the HAND central angle, $\theta_n$. The angles are defined as in Fig. \ref{hall-angles}.}
\center
\begin{tabular}{ccccc}
$\left<Q^2\right>$  & $E_0$  & $\theta_{e'}$ & $p_0$  &  $\theta_n$  \\
(GeV/$c)^2$ & (GeV) & ($^{\circ})$ & (GeV/$c$) & ($^{\circ}$) \\ \hline
0.46 & 2.425 & 17.0 & 2.170 & 62.5 \\
0.96 & 3.605 & 17.0 & 3.070 & 54.0 \\ 
\end{tabular}
\end{table}
 
 \begin{figure}
	\centering
	\includegraphics[width=8cm]{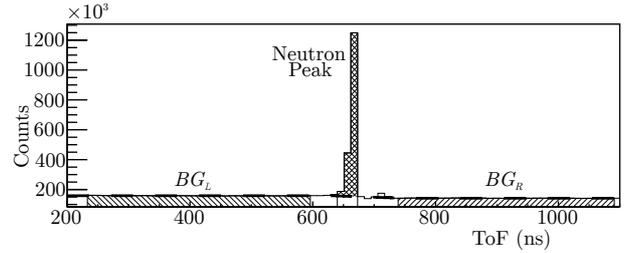}
	\caption [Time of flight spectrum] {Neutrons were identified using the cascading veto method and a time-of-flight spectrum, in this case at $Q^2=0.96~(\mathrm{GeV}/c)^2$. The neutron peak (cross-hatched) was identified using a $\pm 25$~ns cut, and the accidentals background estimated using a linear fit (dashed line) on the background events outside of the cut, shown as $BG_L$ and $BG_R$ (hatched). The small $\gamma$ peak is also visible near 710~ns.}
	\label{tof}
\end{figure}

The kinematics used during this experiment are shown in Table \ref{tab:qekinematics}. 
The number of knocked-out neutrons was determined using the cascading veto method described above along with a 1.5~ms timing window triggered by the scattered electrons. Within that window, a $\pm 25$~ns cut was made on the time-of-flight (TOF) peak to identify neutrons and to separate them from the $\gamma$ peak. Accidentals outside of twice this cut ($\pm 50$~ns) and outside of the nearby $\gamma$ peak were used to estimate and remove background events from the TOF. An example of the TOF spectrum is shown in Fig.~\ref{tof}.
The neutrons were then separated by the orientation of the target spin, identifying those neutrons knocked out when the target spin was `up' as $S_{\uparrow}$ and those when `down' as $S_{\downarrow}$. Each of these was scaled by the respective cumulative beam charge, $C_{\uparrow(\downarrow)}$, and electronic live-time, $L_{\uparrow(\downarrow)}$, to obtain the yields, $Y_{\uparrow (\downarrow)}$, defined by
\begin{equation}
	Y_{\uparrow(\downarrow)} = \frac{S_{\uparrow(\downarrow)}}{C_{\uparrow(\downarrow)}\cdot L_{\uparrow(\downarrow)}}.
\end{equation}
Combined with the polarization of the target, $P_t$, the raw measured asymmetry, $A_{meas}$, was defined as
\begin{equation}
	A_{meas} = \frac{1}{P_t}\left( \frac{Y_{\uparrow}-Y_{\downarrow}}{Y_{\uparrow}+Y_{\downarrow}} \right).
\end{equation}

The true physics asymmetry, $A_y^0$, is scaled due to dilution factors from nitrogen ($D_{\mathrm{N_2}}$) and proton contamination ($D_p$). When these dilutions are taken into consideration, the physics asymmetry takes the form of 
\begin{equation}
	A_y^0 = \frac{A_{meas}}{D_{\mathrm{N}_2}D_{p}}.
\end{equation}

To aid in the SEOP pumping of the $^3$He target, a small amount of N$_2$ was added to inhibit polarization relaxation due to radiation trapping ~\cite{Walker:1997zzc}. The dilution factor from nitrogen contamination was calculated using the pressure curve method \cite{Zhang:2012zzp} and is shown for each kinematic setting in Table \ref{tab:dilution}, where the percentage of nitrogen in the cell is $1-D_{N_2}$.

\begin{table}
\caption{\label{tab:dilution}Contamination due to nitrogen in the target cell ($D_{\mathrm{N_2}}$) and mis-identification of protons ($D_p$) were used to scale the measured asymmetry into the physics asymmetry. The large decrease in $D_{p}$ at the highest $Q^2$ is largely due to the addition of a Pb wall in front of HAND that caused a larger number of mis-identified protons.}
\center
\begin{tabular}{ccccc}
&$\left<Q^2\right>$ (GeV/$c)^2$ & $D_{\mathrm{N}_2}~(\times10^{-2})$ & $D_p~(\times10^{-2})$\\ \hline
&0.46 & $97.9 \pm 0.3$  & $66.7 \pm 0.4$ \\ 
&0.96 & $97.2 \pm 1.2$ & $50.8 \pm 0.5$ \\ 
\end{tabular}
\end{table}

Due to the nature of neutron detection in HAND, there is a possibility that some protons will be misidentified as neutrons, even when utilizing the cascading veto layers method. 
The number of identified neutrons ($N_T$) contains both measured neutrons ($N_n$) and misidentified protons ($N_p$),
\begin{equation}
	N_T = N_n + N_p.
\end{equation}
This relates to the number of neutrons that enter the detector ($n_n$), the neutron detection efficiency ($\epsilon_n$), the number of protons that enter the detector ($n_p$), and the ratio of misidentified protons ($r_{mis}$), 
\begin{equation}	\label{NT}
	N_T = n_n \epsilon_n + n_p r_{mis}.
\end{equation}
$r_{mis}$ was measured by applying neutron cuts on H$(e,e'p)$ scattering and identifying the total number of measured protons ($P_T$) and protons misidentified as neutrons ($P_n$),
\begin{equation}
	r_{mis} = \frac{P_n}{P_T} = \frac{p_n \epsilon_n}{P_T}.
\end{equation}

The number of protons in the detector during $^3$He($e,e'n$) scattering was estimated by calculating the ratio of protons to neutrons ($r_{p:n}$) using the Rosenbluth formula ($\frac{d\sigma}{d\Omega}$) and Kelly fits \cite{Rosenbluth:1950yq, Kelly:2004hm}, 
\begin{equation}
	r_{p:n}=\frac{p_{exp}}{n_{exp}} = \frac{2\left(\frac{d\sigma}{d\Omega}\right)_p}{\left(\frac{d\sigma}{d\Omega}\right)_n},
\end{equation}
and comparing it to the number of neutrons,
\begin{equation}
	n_p = n_n r_{p:n},
\end{equation}
giving
\begin{equation}\label{NTnn}
	N_T = n_n \epsilon_n + n_n r_{p:n} r_{mis}.
\end{equation}
To address having one equation and two unknowns, a tight $x\approx1$ cut and a $\theta:\phi$ cut was made such that the scattered electrons ($E_T$) were both quasi-elastic and sent the nucleons toward HAND ($n_n$ and $n_p$). The same electron cuts were made when HAND was included above in $N_T$ to measure proton dilution. From this, 
\begin{equation}
	E_T = n_n + n_p,
\end{equation}
since no neutron cuts are made and the total electrons counted are scattered from the total knocked-out neutrons and protons. Again, the number of protons are related to the number of neutrons by $n_p = n_n r_{p:n}$, giving
\begin{equation}
	E_T = n_n + n_n r_{p:n}.
\end{equation}
This was re-arranged to get $n_n$ in known terms,
\begin{equation}
	n_n = \frac{E_T}{1+r_{p:n}}.
\end{equation}
Plugging this back into Eq.~\ref{NTnn},
\begin{equation}
	N_T = \left( \frac{E_T}{1+r_{p:n}} \right) \epsilon_n + \left( \frac{E_T}{1+r_{p:n}} \right) r_{p:n} r_{mis},
\end{equation}
where $\epsilon_n$ is determined by 
\begin{equation}
	\epsilon_n = \frac{N_T(1+r_{p:n})}{E_T} - r_{mis} r_{p:n}.
\end{equation}
Values for $\epsilon_n$ are shown in Table~\ref{en} and agreed well with a GEANT4~\cite{Agostinelli:2002hh} simulation that utilized the cascading veto method for neutron identification.

\begin{table}
\caption{\label{en}Neutron detector efficiency determined for each kinematic setting.}

\center
\begin{tabular}{ccc}
$\left<Q^2\right>$ (GeV/$c)^2$ & $\epsilon_n~(\times10^{-2})$ \\ \hline
0.46   &	$5.7$	 \\ 
0.96   &	$28.5$	 \\ 
\end{tabular}
\end{table}

The proton dilution factor is determined by
\begin{equation}
	D_p = \frac{N_n}{N_T} = \frac{n_n \epsilon_n}{N_T},
\end{equation}
\begin{equation}
	D_p = 1-\frac{E_T P_n r_{p:n}}{N_T P_T (1+r_{p:n})}
\end{equation}
with results given in Table~\ref{tab:dilution}.

There were three leading systematic contributions to the total uncertainty of the measurements. The leading contributor was the uncertainty on the target polarization, $\delta P_t$, which was 4.7\%. The uncertainties due to the nitrogen and proton dilutions, $\delta D_{\mathrm{N}_2}$ and $\delta D_{p}$,  are shown in detail in Table \ref{tab:ay-sys}. A summary of all of these contributions to the systematic uncertainty ($\delta A_y^{sys}$) is presented in Table \ref{tab:ay-sys}.

\begin{table}
\caption{\label{tab:ay-sys}Experimental contributions to systematic uncertainty from the target polarization, nitrogen dilution, and proton dilution, each scaled by $10^{-2}$.}
\center
\begin{tabular}{cc|cccc}
& $\left<Q^2\right>$ (GeV/$c)^2$  & 0.46 & 0.96    \\ \hline
& $\delta D_{t}$				& 2.15  & 0.13 \\
& $\delta D_{\mathrm{N}_2}$	 & 0.069 & 0.017 \\
& $\delta D_{p}$	& 0.14 & 0.014 \\ \hline
& $\delta A_y^{sys}$			& 2.15 & 0.13 \\
\end{tabular}
\end{table}

\begin{figure}
	\centering
	\includegraphics{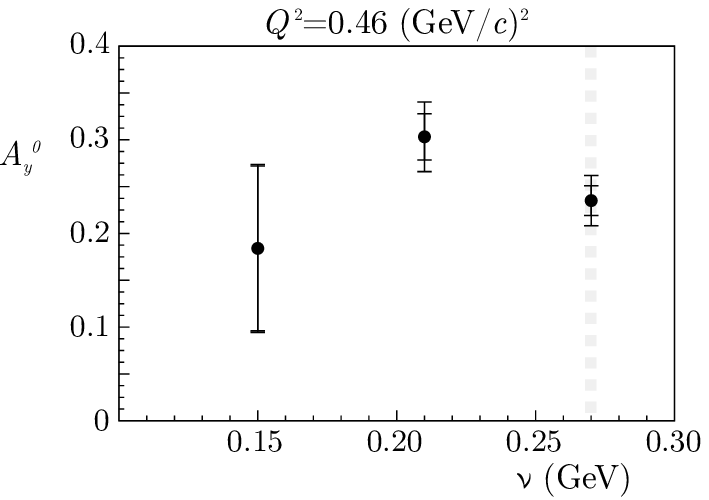} \vspace{10pt}

	\includegraphics{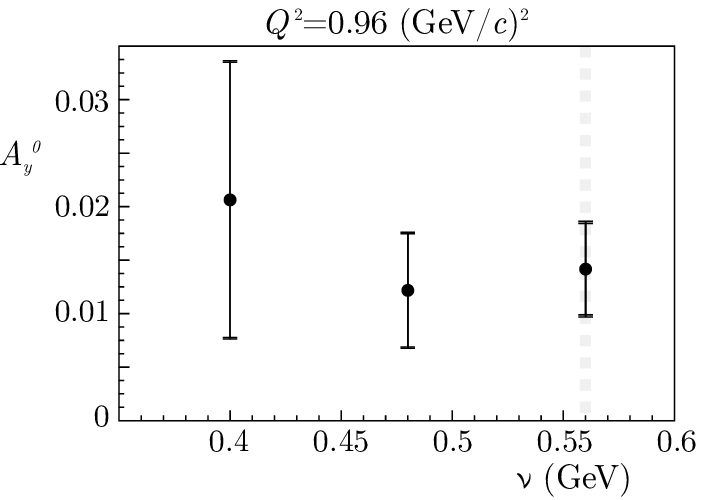}

	\caption [$A_y^0$ vs $\nu$] {Current measurements of $A_y^0$ at $0.46$ and $0.96$~(GeV/$c)^2$ plotted as a function of the energy transfer, $\nu$. The dotted lines indicate the center of the quasi-elastic peak.}
	\label{ay-nu}
\end{figure}

In Fig.~\ref{ay-nu}, $A_y^0$ is plotted as a function of the energy transfer, $\nu$. To further minimize effects from the elastic peak at small $\nu$, only data sitting directly on the quasi-elastic peak were included.

\begin{figure}
	\centering
	\includegraphics[width=8.6cm]{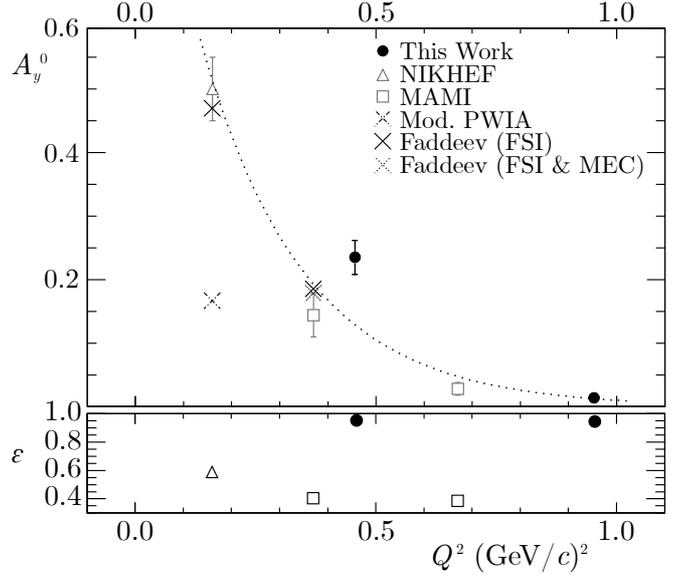}
	\caption [$A_y^0$ World Data] {Current $A_y^0$ measurements, along with the NIKHEF  \cite{Poolman:1999uf} and MAMI \cite{Bermuth:2003qh} data, plotted as a function of $Q^2$ alongside the values of $\varepsilon$ for each data point. Error bars represent the total uncertainties.
	The uncertainties for these data can be found in Table \ref{tab:ay-final}. The dot-dashed cross represents the modified PWIA approach used by Laget \cite{Laget:1991pb,Poolman:1999uf}, the dotted and solid crosses represent the non-relativistic Faddeev calculations including FSI and, in the case of the solid cross, MEC \cite{Bermuth:2003qh}. Only the Faddeev calculations, which fully account for FSI, represent the data. 
The dotted line is an exponential fit of the current world data.}
	\label{ay-final-plot}
\end{figure}

Results from our experiment are presented along with the world data as a function of $Q^2$ in Table \ref{tab:ay-final} and Fig. \ref{ay-final-plot}. As discussed, the non-zero values of $A_y^0$ measured indicates contributions from final state interactions and meson exchange currents. The original Laget calculations \cite{Laget:1991pb,Poolman:1999uf}, calculated using a modified PWIA, greatly underestimated $A_y^0$. Full Faddeev calculations provided by the Bochum group provided reasonable predictions of $A_y^0$ values to both the historical and current data \cite{Bermuth:2003qh}. Faddeev calculations are not available above a $Q^2$ of approximately 0.4~(GeV/$c)^2$ since relativistic effects are not included in the calculations. This experiment is unique in that it reaches unprecedented precision up to $Q^2$ of 0.96~(GeV/$c)^2$, and was also done at much larger $\varepsilon = $ $(1 + 2(1$ $ + $ $ Q^2/4M^2)\tan^2\theta_e/2)^{-1}$ than previous results, a region that has been shown to be sensitive to effects beyond the Born approximation such as two-photon exchange~\cite{Meziane:2010xc,Zhang:2015kna}.
$A_y^0$ is large at low $Q^2$, where FSI and MEC are significant, and drops off exponentially to the $10^{-2}$ level as $Q^2$ approaches 1~(GeV/$c)^2$, where contributions from FSI and MEC are greatly reduced. Any extractions of the neutron's electromagnetic form factors from $^3$He scattering must account for these effects.

\begin{table}
\caption{\label{tab:ay-final}Experimental results for $A_y^0$ scaled by $10^{-2}$. }
\center
\begin{tabular}{cccc}
& $\left<Q^2\right>$ (GeV/$c)^2$	 & $A_y^0 \pm\delta A_y^{stat}\pm\delta A_y^{sys}$ \\ \hline
&	0.46			& $	23.5	\pm 1.58	\pm	2.15	$ \\
&	0.96			& $	1.42 	\pm	0.43	\pm	0.13	$ \\
\end{tabular}
\end{table}

We thank the Jefferson Lab Hall~A technical staff and the Jefferson Lab accelerator staff for their outstanding support. This work was supported in part by the National Science Foundation, the U.S. Department of Energy, and the UK Science and Technology Facilities Council. Jefferson Science Associates, LLC, operates Jefferson Lab for the U.S. DOE under U.S. DOE contract DE-AC05-06OR23177.

\bibliography{bibliography}			

\end{document}

%% file: author_list_plb.tex

\author[UNH]{E.~Long\corref{ca}}
\cortext[ca]{Corresponding author}
\ead{elena.long@unh.edu}

\author[Rutgers]{Y.W.~Zhang} 
\author[JSI]{M.~Mihovilovi\v{c}}
\author[UVA]{G.~Jin} 
\author[MIT]{V.~Sulkosky}
\author[MIT]{A.~Kelleher}
\author[KentState]{B.~Anderson} 
\author[JLab]{D.W.~Higinbotham} 
\author[JSI]{S.~\v{S}irca}
\author[JLab]{K.~Allada}
\author[Glasgow]{J.R.M.~Annand} 
\author[WM]{T.~Averett} 
\author[MIT]{W.~Bertozzi}
\author[FIU]{W.~Boeglin}
\author[WM]{P.~Bradshaw}
\author[JLab]{A.~Camsonne}
\author[ODU]{M.~Canan} 
\author[UVA]{G.D.~Cates}
\author[Hampton]{C.~Chen} 
\author[JLab]{J.-P.~Chen}
\author[JLab]{E.~Chudakov}
\author[Bari]{R.~De~Leo} 
\author[UVA]{X.~Deng}
\author[JLab]{A.~Deur}
\author[UKentucky]{C.~Dutta} 
\author[Rutgers]{L.~El~Fassi}
\author[Temple]{D.~Flay}
\author[INFN]{S.~Frullani} 
\author[INFN]{F.~Garibaldi}
\author[Duke]{H.~Gao} 
\author[MIT]{S.~Gilad}
\author[JLab]{R.~Gilman}
\author[Kharkov]{O.~Glamazdin} 
\author[ODU]{S.~Golge}
\author[JLab]{J.~Gomez}
\author[JLab]{J.-O.~Hansen}
\author[Longwood]{T.~Holmstrom} 
\author[MIT,LosAlamos]{J.~Huang}
\author[Cairo]{H.~Ibrahim}
\author[JLab]{C.W.~de~Jager}
\author[CNU]{E.~Jensen}
\author[LosAlamos]{X.~Jiang}
\author[JLab]{M.~Jones}
\author[Seoul]{H.~Kang}
\author[WM]{J.~Katich}
\author[FIU]{H.P.~Khanal}
\author[OhioU]{P.M.~King}
\author[UKentucky]{W.~Korsch}
\author[JLab]{J.~LeRose}
\author[UVA]{R.~Lindgren}
\author[Huangshan]{H.-J.~Lu}
\author[Lanzhou]{W.~Luo} 
\author[FIU]{P.~Markowitz}
\author[WM]{M.~Meziane}
\author[JLab]{R.~Michaels}
\author[JLab]{B.~Moffit}
\author[Hampton]{P.~Monaghan}
\author[MIT]{N.~Muangma}
\author[JLab]{S.~Nanda}
\author[UVA]{B.E.~Norum}
\author[MIT]{K.~Pan}
\author[CMU]{D.~Parno}
\author[TelAviv]{E.~Piasetzky}
\author[Temple]{M.~Posik}
\author[NSU]{V.~Punjabi} 
\author[MIT,LosAlamos]{A.J.R.~Puckett}
\author[Duke]{X.~Qian}
\author[JLab]{Y.~Qiang}
\author[Lanzhou]{X.~Qui}
\author[UVA,MIT]{S.~Riordan}
\author[JLab,Deceased]{A.~Saha}
\author[JLab]{B.~Sawatzky}
\author[UVA]{M.~Shabestari}
\author[YPI]{A.~Shahinyan}
\author[NorthMU]{B.~Shoenrock}
\author[Longwood]{J.~St.~John}
\author[GWU]{R.~Subedi}
\author[UVA]{W.A.~Tobias}
\author[NorthMU]{W.~Tireman}
\author[INFN]{G.M.~Urciuoli}
\author[UVA]{D.~Wang}
\author[UVA]{K.~Wang}
\author[UIUrbCh]{Y.~Wang}
\author[KentState]{J.~Watson}
\author[JLab]{B.~Wojtsekhowski}
\author[Hampton]{Z.~Ye}
\author[MIT]{X.~Zhan}
\author[Lanzhou]{Y.~Zhang}
\author[UVA]{X.~Zheng}
\author[WM]{B.~Zhao}
\author[Hampton]{L.~Zhu}


\address[UNH]{University of New Hampshire, Durham, NH, 03824, USA} 
\address[Rutgers]{Rutgers University, New Brunswick, NJ, 08901, USA}
\address[JSI]{Jo\v{z}ef Stefan Institute, Ljubljana 1000, Slovenia}
\address[UVA]{University of Virginia, Charlottesville, VA, 22908, USA}
\address[MIT]{Massachusetts Institute of Technology, Cambridge, MA, 02139, USA}
\address[KentState]{Kent State University, Kent, OH, 44242, USA}
\address[JLab]{Thomas Jefferson National Accelerator Facility, Newport News, VA 23606, USA}
\address[Glasgow]{Glasgow University, Glasgow, G12 8QQ, Scotland, United Kingdom}
\address[WM]{The College of William and Mary, Williamsburg, VA, 23187, USA}
\address[FIU]{Florida International University, Miami, FL, 33181, USA}
\address[ODU]{Old Dominion University, Norfolk, VA, 23508, USA}
\address[Hampton]{Hampton University , Hampton, VA, 23669, USA}
\address[Bari]{Universite di Bari, Bari, 70121 Italy}
\address[UKentucky]{University of Kentucky, Lexington, KY, 40506, USA}
\address[Temple]{Temple University, Philadelphia, PA, 19122, USA}
\address[INFN]{Istituto Nazionale Di Fisica Nucleare, INFN/Sanita, Roma, Italy}
\address[Duke]{Duke University, Durham, NC, 27708, USA}
\address[Kharkov]{Kharkov Institute of Physics and Technology, Kharkov 61108, Ukraine}
\address[Longwood]{Longwood University, Farmville, VA, 23909, USA}
\address[LosAlamos]{Los Alamos National Laboratory, Los Alamos, NM, 87545, USA}
\address[Cairo]{Cairo University, Cairo, Giza 12613, Egypt}
\address[CNU]{Christopher Newport University, Newport News, VA, 23606, USA}
\address[Seoul]{Seoul National University, Seoul, Korea}
\address[OhioU]{Ohio University, Athens, OH, 45701, USA}
\address[Huangshan]{Huangshan University, People's Republic of China}
\address[Lanzhou]{Lanzhou University, Lanzhou, Gansu, 730000, People's Republic of China}
\address[CMU]{Carnegie Mellon University, Pittsburgh, PA, 15213, USA}
\address[TelAviv]{Tel Aviv University, Tel Aviv 69978, Israel}
\address[NSU]{Norfolk State University, Norfolk, VA, 23504, USA}
\address[YPI]{Yerevan Physics Institute, Yerevan, Armenia}
\address[NorthMU]{Northern Michigan University, Marquette, MI, 49855, USA}
\address[GWU]{George Washington University, Washington, D.C., 20052, USA}
\address[UIUrbCh]{University of Illinois at Urbana-Champaign, Urbana, IL, 61801, USA}

\fntext[deceased]{Deceased 9 May 2011.}